\documentclass[12pt,preprint]{aastex}
\usepackage{epsfig}
\usepackage{subfig}

\newcommand{\msun}{\,\rm M_\odot}

\newcommand{\be}{\begin{equation}}
\newcommand{\ee}{\end{equation}}
\newcommand{\ba}{\begin{eqnarray}}
\newcommand{\ea}{\end{eqnarray}}

\newcommand{\rtwo}{r_{\rm 200}}

\newcommand{\Vmax}{V_{\rm max}}
\newcommand{\rVmax}{r_{\rm Vmax}}

\newcommand{\mcut}{m_0}
\newcommand{\sigv}{\langle \sigma v \rangle}
\newcommand{\cmcps}{{\rm cm}^3\;{\rm s}^{-1}}

\begin{document}

\title{Fermi-LAT Sensitivity to Dark Matter Annihilation in Via~Lactea~II
Substructure}

\author{Brandon Anderson$^1$, Michael Kuhlen$^{2}$, J\"urg Diemand$^{3,4}$,
Robert Johnson$^1$, \& Piero Madau$^{1,3}$}

\altaffiltext{1}{ Santa Cruz Institute for Particle Physics, University of
California Santa Cruz, 1156 High St., Santa Cruz CA 95064}
\altaffiltext{2}{Department of Astronomy, University of California Berkeley, 501
Campbell Hall, Berkeley CA 94709}
\altaffiltext{3}{Department of Astronomy \& Astrophysics, University of
California Santa Cruz, 1156 High St., Santa Cruz CA 95064}
\altaffiltext{4}{Institute for Theoretical Physics, University of Zurich,
Winterthurerstr. 190, 8057 Zurich, Switzerland}
\email{anderson@physics.ucsc.edu, mqk@astro.berkeley.edu}

\begin{abstract}
We present a study of the ability of the Fermi Gamma-ray Space Telescope to
detect dark-matter annihilation signals from the Galactic subhalos predicted by
the Via Lactea~II N-body simulation.
We implement an improved formalism for estimating the boost factor needed to
account for the effect of dark-matter clumping on scales below the resolution of
the simulation, and we incorporate a detailed Monte Carlo simulation of the
response of the Fermi-LAT telescope, including a simulation of its all-sky
observing mode integrated over a ten year mission.
We find that for WIMP masses up to about 150~GeV/$c^2$ in standard
supersymmetric models with $\langle \sigma v\rangle=3\times 10^{-26} {\rm
cm}^{3} {\rm s}^{-1}$, a few subhalos could be detectable with $>5$ standard
deviations significance and would likely deviate significantly from the
appearance of a point source.
\end{abstract}


\section{Introduction}

Even 75 years after the first observational evidence for a
non-luminous form of matter \citep{Zwicky1933} we know remarkably
little about the nature of the hypothesized dark matter particle. A
promising way forward is indirect detection, whereby ground and
space-based observatories are searching for the products of pair
annihilations of dark matter particles, such as neutrinos,
relativistic positrons, or gamma-rays.

The gamma-ray signal in particular has received a lot of attention in
recent years, owing in part to the launch of the Fermi Gamma-ray Space
Telescope. Numerous papers have discussed whether the Large Area
Telescope (LAT) aboard Fermi will be sensitive enough to detect this
signal and how to differentiate it from conventional
astrophysical sources, e.g. \citet{Baltz2007}. It appears that a detection
is challenging but feasible for a wide range of plausible physics models
\citep{Baltz2008}. The Galactic Center (GC) has the greatest and closest
concentration
of dark matter in the Local Group and as such is a
promising target \citep{Berezinsky1994, Bergstroem1998, Cesarini2004,
  Jeltema2008}. It is, unfortunately, also a very active region,
with a bright diffuse flux of gamma-rays from cosmic-ray interactions as well as
being
filled with numerous hard X-ray and gamma-ray sources such as supernova
remnants, pulsar wind nebulae, X-ray binaries, etc.
\citep{Aharonian2006, Kuulkers2007}, which complicate attempts 
to search for a DM annihilation signal.
Alternatively, the diffuse gamma-ray signal from the
Milky Way host halo out beyond several degrees from the GC has been
suggested as the most promising \citep{Stoehr2003,
  Springel2008}, but it will be difficult to disentangle from the
poorly constrained diffuse Galactic gamma-ray background arising from cosmic-ray
interactions with interstellar hydrogen and the interstellar radiation
field \citep{Strong2004}.
Instead, the centers of
Galactic subhalos may prove to be the most detectable and least ambiguous
sources of gamma-rays
from DM annihilations. These could be either subhalos hosting dwarf
satellites \citep{Strigari2008} or one of the many dark subhalos
predicted by numerical simulations \citep[e.g.][]{Diemand2008, Kuhlen2009}.

Subhalos as sources have recently been considered by
\citet{Kuhlen2008}, who used the \textit{Via Lactea II} simulation
(VL2) in conjunction with a realistic treatment of the expected
backgrounds to show that a handful, even up to a few dozen, subhalos
should be detectable with the Fermi-LAT at more than $5\sigma$ significance
for a standard weakly interacting DM particle with a mass between 50
and 500~GeV and a cross section in the range $\sigv \sim 10^{-26} -
10^{-25} \cmcps$. That analysis assumed homogenous sky-coverage, an angular
resolution of 9$^\prime$, and used a pre-launch
estimate of the energy-dependent effective area of the LAT. The present paper
updates and improves upon that analysis by running the same VL2 all-sky
emission maps and the diffuse background predictions through a Monte Carlo
simulation of the Fermi-LAT instrument,
taking into account the time-dependent sky coverage and an
energy-dependent instrument response and angular
resolution. Additionally we introduce an improved treatment of the
boost factor due to the DM substructure below the resolution limit of the N-body
simulation.

\section{Methods}
\label{sec_methods}

The VL2 simulation is a one of the highest resolution
cosmological simulations of the formation of a Milky-Way-scale dark
matter halo to date \citep{Diemand2008}.
It employs just
over one billion $4,100 \msun$ particles to model the formation of a
$M_{200}=1.93\times 10^{12}\,\msun$ Milky-Way size halo and its
substructure. It resolves over 50,000 subhalos today within the host's
$\rtwo=402$~kpc (the radius enclosing an average density 200 times the 
cosmological mean matter density). As described in more detail in \citet{Kuhlen2008},
the simulated dark matter distribution was used to construct all-sky
maps of the annihilation flux for a set of observers located 8~kpc
from the host halo's center. These maps consist of $2400 \times 1200$
pixels equally spaced in longitude and the cosine of colatitude,
corresponding to a solid angle per pixel of $4.4 \times 10^{-6}$~sr,
and form the basis for our studies assessing the capability of the Fermi-LAT to
detect
signals from individual subhalos. To correct
for the artificially low central densities of poorly resolved
subhalos, the surface brightness from the central region
of each subhalo was increased on a pixel-by-pixel basis to
match the expected surface brightness of an NFW~\citep{NFW1996} halo with the
subhalo's
measured $\Vmax$ and $\rVmax$. Note that assuming an NFW profile is somewhat
conservative: using the density profiles measured directly in large
N-body simulations (e.g.\ VL2, Springel et al 2008) or one of
the newer fitting functions instead of NFW leads to an increase of
about 30\% in the halo luminosity \citep{DiemandMoore2010}.

In an attempt to account for the increase in luminosity due to
clumping of the dark matter distribution below the scales resolved by the
simulation, we ``boost'' the total flux from a subhalo of mass $M$
by a factor $B(M)$, determined from an analytic model
\citep[described in][]{Kuhlen2008} that depends on the slope $\alpha$
and low mass cutoff $\mcut$ of the subhalo mass function. In \citet{Kuhlen2008}
this
resulted in a factor of $\sim 2$ difference in the number of
detectable subhalos depending on the values of these uncertain
parameters. The prescription applied there, however,
does a poor job of accounting for the expected radial distribution of
this boost. By summing $(1+B(M)) \rho_i m_i$ (where $m_i$ and $\rho_i$ are the mass and density of the $i^{\rm th}$
simulation particle) over all
the subhalo's particles within a given pixel, the radial dependence of
the boost effectively followed that of the subhalo's smooth
\textit{luminosity} profile, whereas it really should have followed
the radial profile of the subhalo population. As such it overly boosted the
bright central
region, and the results for the most strongly boosted scenarios
($\alpha=2.0$) were overly optimistic.

\begin{figure}[h]
\centerline{\psfig{file=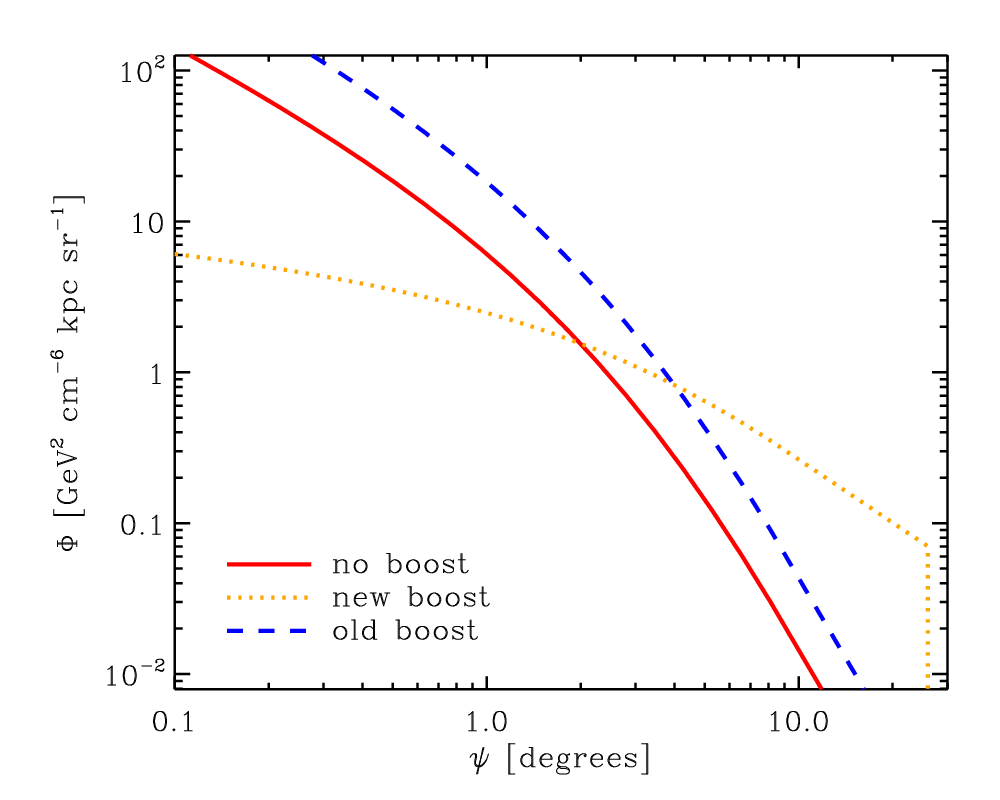,height=3in}}
\caption{A comparison of the old and new boost prescriptions, performed before the
observation simulation on the
prominent sub
halo near the right edge of the projection in Fig.~\ref{fig_allsky:subfig4}.  The dotted
line represents the new boost; the dashed and solid represent the old and un-boosted
profiles, respectively.}
\label{fig_analytical_profile}
\end{figure}

In this work we apply the boost factor in a way that more
appropriately accounts for the radial distribution of
subhalos.\footnote{The numerical simulations do not have sufficient resolution 
to address directly the radial distribution of sub-subhalos within subhalos, 
so we are guided by the radial distribution of subhalos within the host halo.}
The boosted luminosity of a subhalo of mass $M$ is now given
by
\begin{equation}
L = \sum m_i \left( \rho_i + B(M)\langle \rho \rangle \right).
\end{equation}
The sum is over all of the subhalo's particles, and
$\langle
\rho \rangle = (\sum m_i \rho_i)/M$ is the particle-mass-weighted mean
density of the subhalo. Note that just as in the original
prescription, the total subhalo luminosity equals $(1+B(M))$ times the
smooth luminosity, but that the radial dependence of the boosted
component (i.e. $m_i\;B(M)\langle \rho \rangle$) now follows the
\textit{mass}, not the luminosity. See Figure~\ref{fig_analytical_profile} for a comparison
of the different boost intensity profiles.  This new prescription may
still lead to an overly centrally concentrated boost, since the radial
number density profile of subhalos is known to be anti-biased with
respect to the host's mass density profile~\citep{Diemand2007}. 
Note however that the size
of this anti-bias depends strongly on the way a subhalo sample is
defined.
Springel et al.\ 2008 used samples selected by the present
subhalo mass, which show the largest anti-bias, but are irrelevant for
the luminosity distribution of substructure. Tidal mass loss affects
subhalo masses much more than subhalo luminosities. The mass of
subhalos near the halo center is reduced the most,
introducing a strong anti-bias into any mass selected sample. But the samples
relevant
here consist of subhalos that, despite the tidal effects, remain
above a certain luminosity, not above a certain mass. Such luminosity
selected subhalo samples show practically no anti-bias.  Only within
about 5\% of the virial radius does the subhalo number density profile
become shallower than the mass density profile (Figure~6 in \citet{DiemandMoore2010}). 
There, $\rho(r) > 200 \bar{\rho} > 20 B(M) \bar{\rho}$.
In other words, Eqn.~(1) follows the real luminosity distribution very well
and the correction in the very inner parts over-estimates the
total luminosity by a few percent at most.  We use a boost with
$(\alpha,m_{0})=(2.0,10^{-6}M_{\odot})$, normalized to have 10\% of the mass in
clumps containing between 10$^{-5}$ and 10$^{-2}$ times the total mass.
This provides an upper bound for the boost factor,
and we include the overly pessimistic unboosted case as a lower bound.

\subsection{Observation Simulation}
The Fermi-LAT observation simulation program, \emph{gtobssim}, that is part of
the LAT
Science Tools package supported by the Fermi Science Support Center (FSSC), uses
parameterized instrument response functions (based on detailed Monte-Carlo
simulations backed up by beam testing) to approximate the response of the LAT
instrument in orbit.\footnote{\underline{http://Fermi.gsfc.nasa.gov/ssc}}  
For each source, the user provides the spacecraft pointing history, a FITS
sky map of the sources, and the total photon flux.  Simulating operation in
``sky-survey'' mode (nearly uniform exposure) over a ten year period, we
generate
realistic diffuse background predictions as well as predictions for the
WIMP (Weakly Interacting Massive Particle) gamma-ray signal.

The simulation includes the dependence of the LAT effective area on the viewing
angle and photon energy, after accounting for all selection effects, including
the trigger, the on-board filter, and the extensive offline analysis used to
reduce cosmic-ray background to a low level.
The simulation of the point-spread function (PSF) accounts for the
dependencies on inclination angle and energy and also includes a parametrization
of
the significant non-Gaussian tails.
Conversions in the thin versus thick tungsten foils are separately
parameterized, an important detail considering that almost half of the LAT
effective area is from thick-foil conversions, for which the angular resolution
is roughly a factor of two worse.
The observation simulation includes the rocking of the instrument
toward the orbital poles to improve uniformity of the all-sky exposure, and it
also takes into account dead time, in particular the passages through the South
Atlantic Anomaly, during which triggering of the LAT is
disabled~\citep{Atwood2009}.

We consider WIMPs that are capable of pair
annihilation, for example a hypothetical stable supersymmetric partner of the
gauge and Higgs bosons, namely the neutralino.
The WIMP remains a promising candidate for the role of DM, in particular for
indirect detection by astronomical observations \citep{Bergstroem2000}.
After choosing a mass, $M_{\chi}$, and a thermally-averaged, velocity-weighted
annihilation cross section, $\langle \sigma v\rangle$, we use
DarkSUSY~\citep{Gondolo2004} to simulate by Monte-Carlo the gamma ray spectrum
that results from WIMP annihilation and the subsequent fragmentation.
In keeping with the model used in \citet{Kuhlen2008}, we compute this spectrum 
for WIMP masses ranging from 50 to
500~GeV, assuming a 100\% branching ratio into $b\bar{b}$ quarks and $\langle
\sigma v\rangle=3\times 10^{-26} {\rm cm}^{3} {\rm s}^{-1}$.  Fig.~\ref{fig_spectral_breakdown} 
shows how a 100~GeV spectrum compares with the backgrounds in different regions of the sky.

Since the energy spectrum of the DM signal has no correlation with the spatial
distribution, we calculate the overall flux for input to \emph{gtobssim} as
\begin{equation}
\Phi=\frac{1}{4\pi} \frac{\langle \sigma v\rangle}{2 M_{\chi}^{2}}
\int_{E_{TH}}^{M_{\chi}}
\frac{dN}{dE} dE \int_{all~sky} d \Omega\int_{los} \rho(l)^{2}~dl ,
\end{equation}
where $\rho(l)$ is the density in the VL2 simulation, $dN/dE$ is the
annihilation spectrum, and $E_{TH}=500$~MeV is the minimum photon energy
accepted in our analysis.

\begin{figure}[ht]
\subfloat[DM Host Halo]{
\includegraphics[scale=0.2]{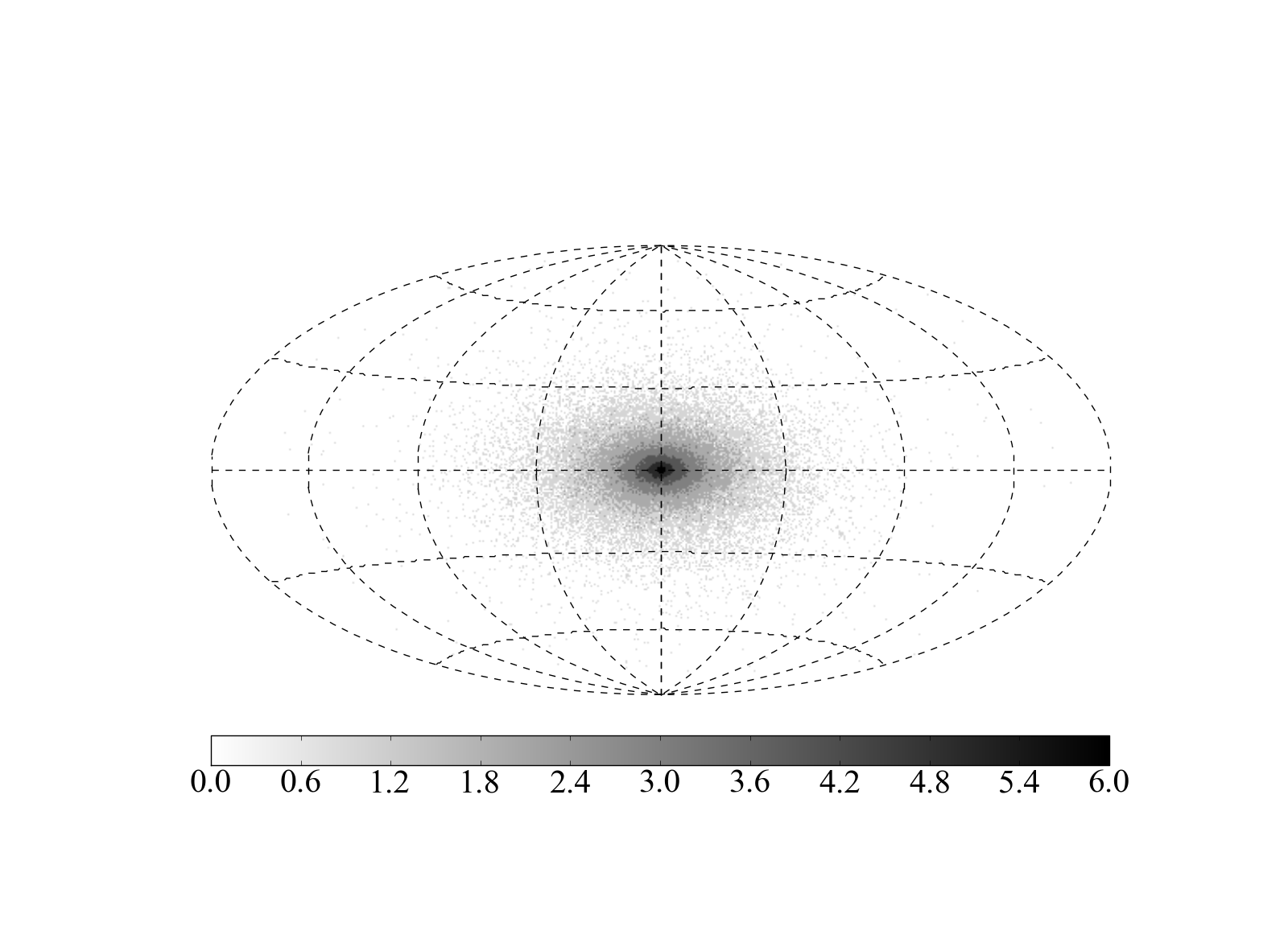}
\label{fig_allsky:subfig1}
}
\subfloat[DM Host Halo + DM Diffuse]{
\includegraphics[scale=0.2]{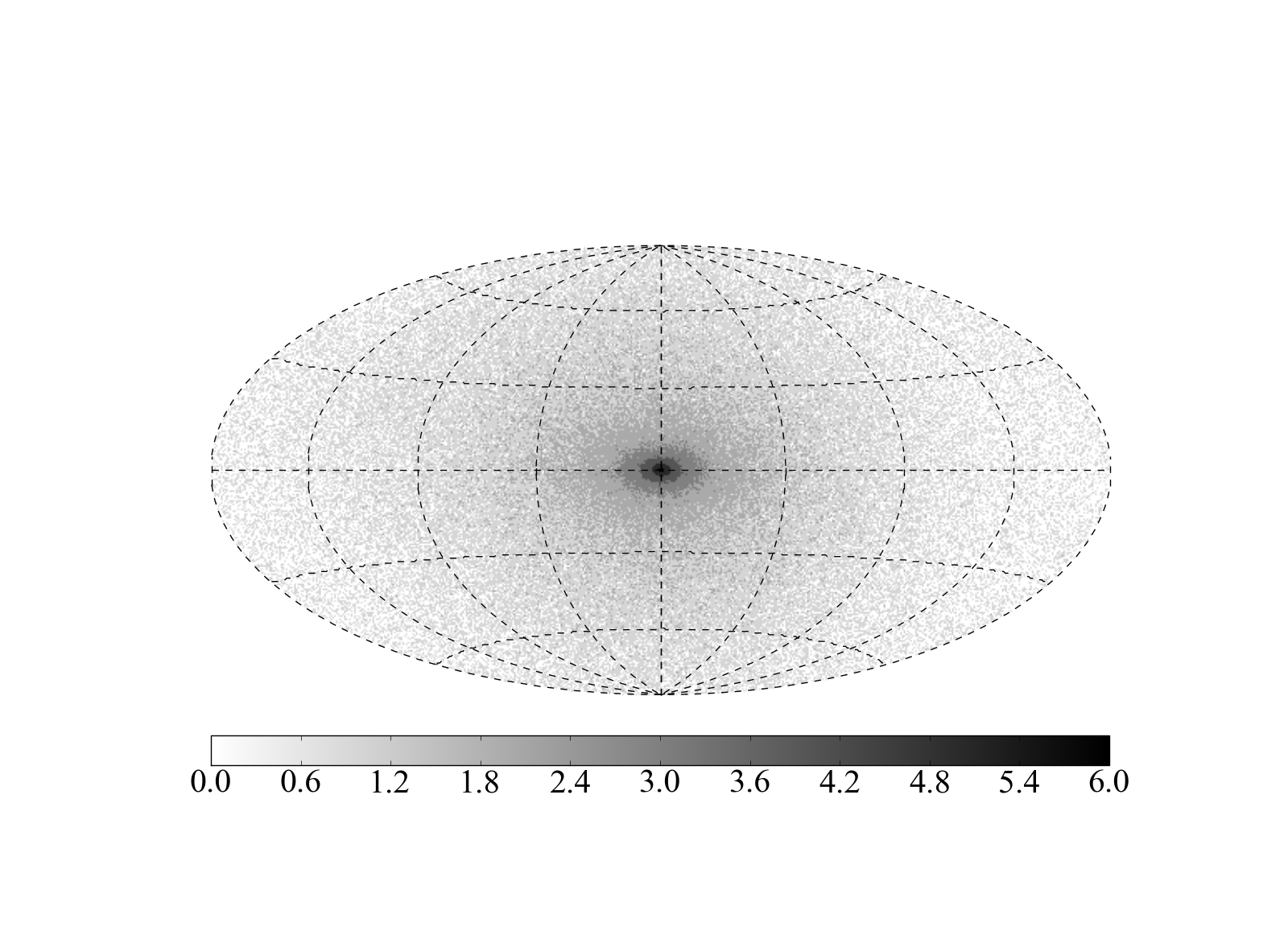}
\label{fig_allsky:subfig2}
}
\\
\subfloat[Galactic + Extragalactic]{
\includegraphics[scale=0.2]{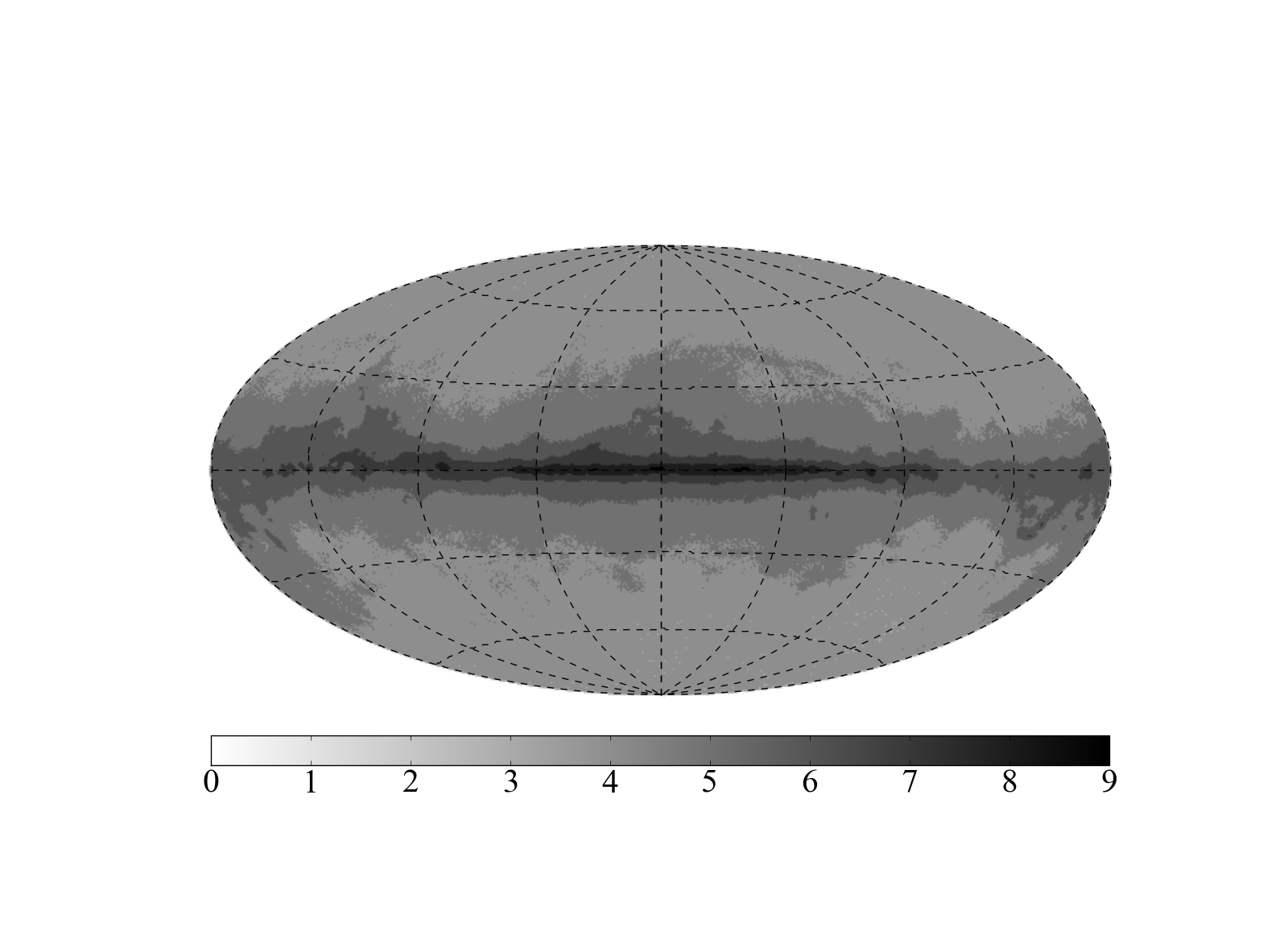}
\label{fig_allsky:subfig3}
}
\subfloat[Resolved Subhalos]{
\includegraphics[scale=0.2]{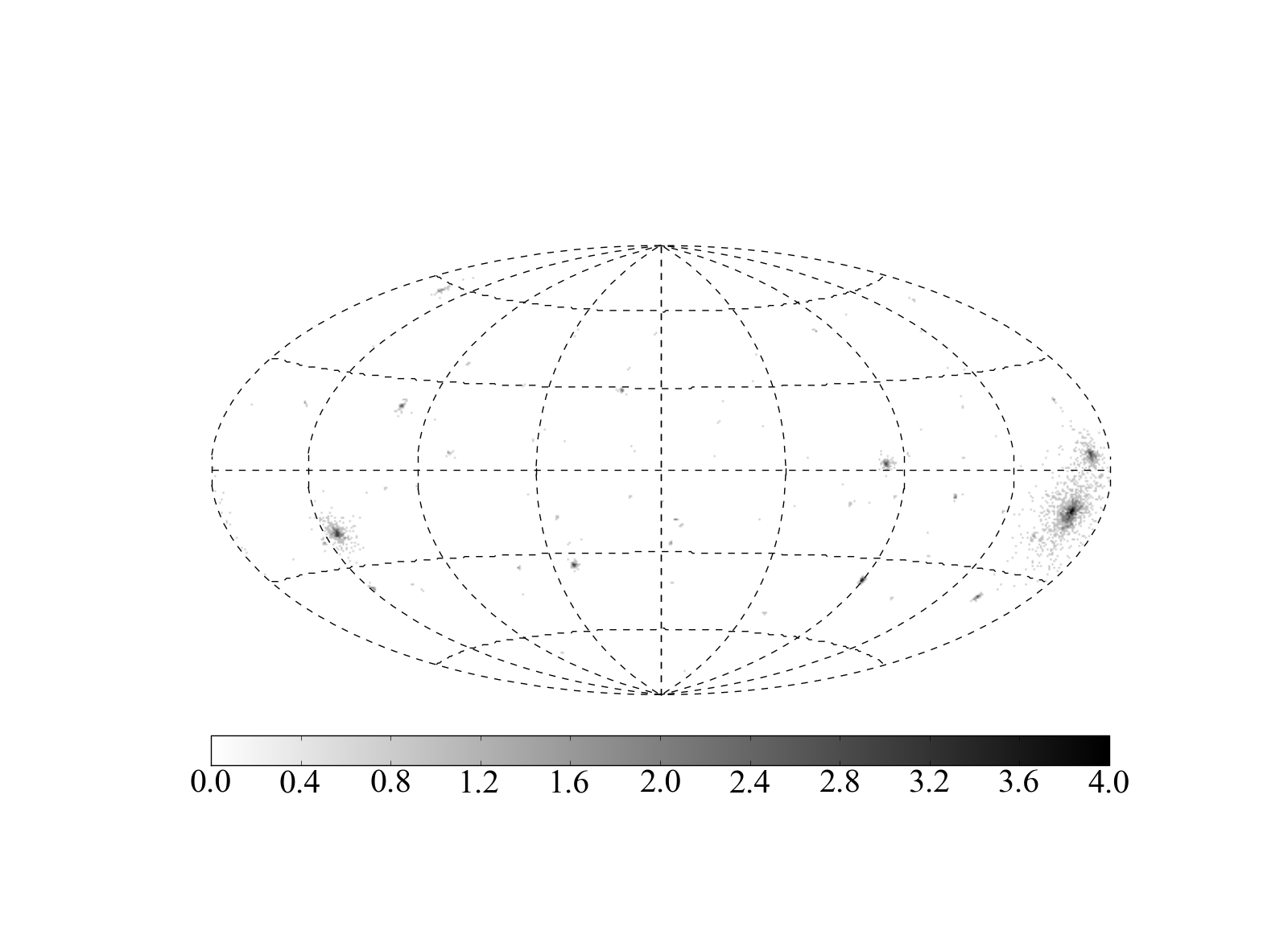}
\label{fig_allsky:subfig4}
}
\caption{All-sky $log(\rm{counts})$ maps for 100~GeV WIMP subhalo annihilation signal 
and the three background sources for ten years of Fermi-LAT orbit 
as seen by an observer along the intermediate halo axis.  The second panel 
(\ref{fig_allsky:subfig2}) includes an extrapolation for subhalos which are unresolved
in the VL2 simulation, which we call DM Diffuse.
}
\label{fig_allsky}
\end{figure}

\begin{figure}[ht]
\subfloat[All-sky]{
\includegraphics[scale=0.2]{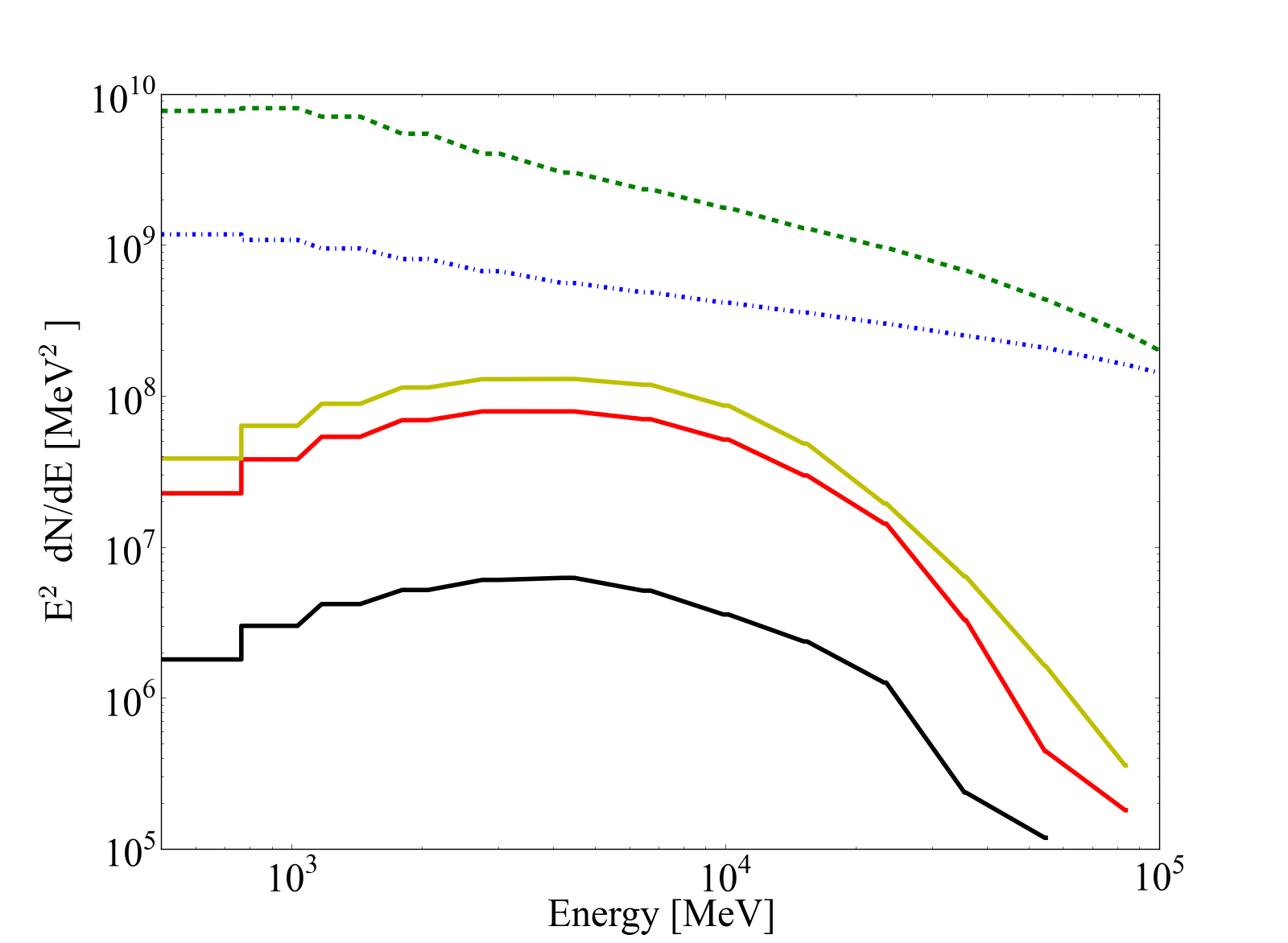}
\label{fig_spectral_breakdown:subfig1}
}
\subfloat[Single Subhalo]{
\includegraphics[scale=0.2]{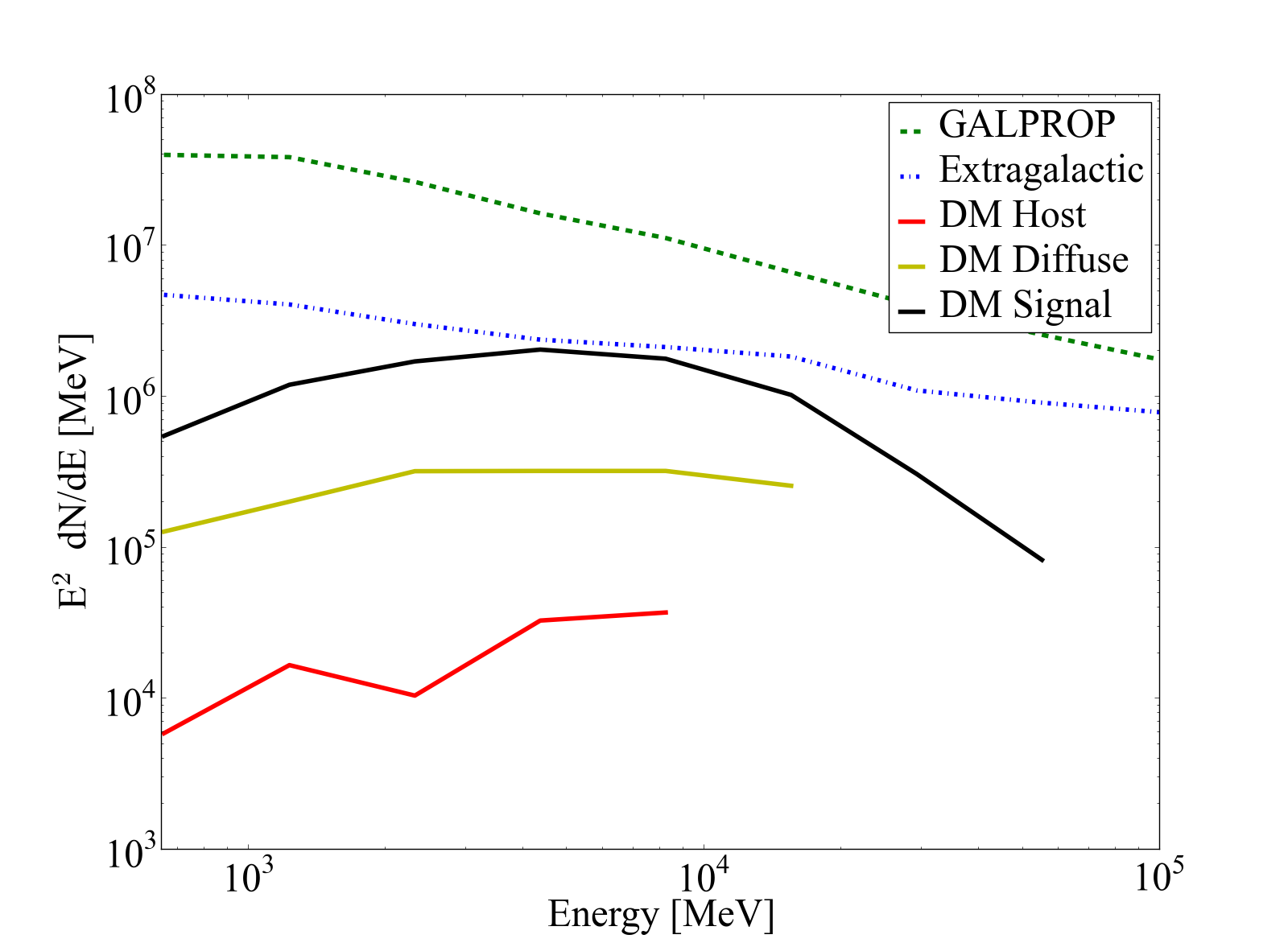}
\label{fig_spectral_breakdown:subfig2}
}
\caption{The relative contributions to the entire sky, including all subhalos (\ref{fig_spectral_breakdown:subfig1}), 
and the inner five degrees of a single detectable subhalo (\ref{fig_spectral_breakdown:subfig2}) 
for ten years of simulated Fermi data from a 100 GeV WIMP and the four background 
sources.  The single subhalo is located at (l, b) = (195.45, -11.25).  Bins are 
equally spaced in log(E).}  
\label{fig_spectral_breakdown}
\end{figure}

We repeated the simulations for each WIMP mass using VL2 subhalo sky maps
corresponding to 10 random viewpoints at 8~kpc radius around the VL2 galaxy.
A total of ten years of Fermi-LAT operation was simulated for each viewpoint.
Figure~\ref{fig_allsky} shows the full sky for an example simulation of the
signal over a ten year observation.

The DM subhalo signal competes with a diffuse background from four categories:
extragalactic, Galactic, host-halo DM, and unresolved DM.  We simulate
extragalactic 
diffuse emission using the power law fit determined in the recent Fermi-LAT
analysis~\citep{extragalactic}.
For Galactic emission, we use the GALPROP~\citep{Strong2000} (v50.1p) cosmic-ray
propagation code to produce a map of diffuse gamma-rays originating from cosmic
ray interactions with the Galactic interstellar medium and radiation field.
The ``optimized''GALPROP model \citep{Strong2004} used in \citet{Kuhlen2008} 
relaxes constraints imposed by measurements of the local cosmic-ray flux, especially 
for cosmic-ray electrons, in order to account for the ``GeV excess'' seen by EGRET \citep{Hunter1997}.
Since initial Fermi-LAT observations show no such
excess~\citep{Johannesson2009}, 
we return to the ``conventional'' model, which keeps the constraints imposed by
local cosmic-ray fluxes.
We did not include the LAT residual cosmic-ray background in the model.
It is roughly one third the extragalactic diffuse contribution and if included
would
decrease our sensitivity to DM subhalo objects by an estimated 20\%.

Galactic sources were not included in the background model.
Since their gamma-ray signals are likely to overlap with subhalos,
especially at low energy, they will reduce the sensitivity with 
respect to what is presented here.
Furthermore, two or more point sources that are nearly coincident on the sky can
mimic an
extended object in the gamma-ray signal, producing a background to be dealt with
when analyzing Fermi-LAT data.
In general, when a subhalo candidate is detected it will be necessary to try to
reject point
source hypotheses through spectral analysis, analysis of the angular shape or
source extension (see \S~\ref{secAngular}), searches for temporal variations,
and multi-wavelength studies \citep{Baltz2007}.

In addition to the Galactic and extragalactic diffuse, the background includes
photons from annihilation of DM residing in the smooth host halo.  For the case
where we boost the substructure, this smooth component also includes the extrapolated
contribution due to unresolved subhalos.  The total mass of the extrapolated substructure
has the same normalization as the boost described in \S~\ref{sec_methods}, but 
an anti-biased distribution due to tidal stripping and destruction of subhalos at inner
radii.  \citep{Kuhlen2008}

We generated ten-year Fermi-LAT Monte Carlo simulations of all four diffuse
background sources using \emph{gtobssim}, by the same procedure as used for the
signal
subhalos.  All Monte-Carlo simulation counts were binned, by the Fermi-LAT
Science Tool \emph{gtbin}, into four logarithmically spaced energy bands from
500~MeV to 300 GeV for use in \S~\ref{secAngular}.

In order to assign a detection significance to the region occupied by a
particular subhalo, we first find the total simulated background counts
$\lambda$ in the region
and then calculate the Poisson probability
\begin{equation}
P = \displaystyle\sum_{i=k}^{\infty} \frac{\lambda^{i} e^{-\lambda}}{i!}
\end{equation}
for the background to fluctuate to a level equal to or greater than the observed
counts $k$.
From $P$ we derive a significance expressed in terms of standard deviations for
a normal distribution:
\begin{equation}
S = \sqrt{2}\rm{~erf}^{-1}(1-2\emph{P}).
\end{equation}

For each subhalo we take all photons above 500~MeV in a region of interest (ROI) 
set at one degree, which corresponds roughly to the 68\%
containment angle in the LAT at 500 MeV.  ~\citep{Atwood2009} Fine-tuning this radius does not
dramatically change the results given here.  Finally, for a given map we quote $N_{5}$
($N_{3}$), the number of subhalos above five (three) standard deviations significance.

\subsection{Resolution of Angular Structure}
\label{secAngular}
One of the crosschecks needed to bolster the case for a DM subhalo candidate is
to look for extended emission that is inconsistent with a point source.  Since
the LAT PSF has a strong energy dependence, we employ a likelihood-ratio
analysis to compare for each significant subhalo the point-source hypothesis to
a subhalo hypothesis.  Instead of using a spatially unbinned event list, as in
the previous section, we bin the sky into a grid of 2400 $\times$ 1200 pixels,
equally spaced in Galactic longitude and latitude.  For candidates near the
galactic poles the counts are binned with a coordinate-system rotated 90 degrees 
in l and b to avoid distortion.  Then for each VL2 subhalo,
we convolve the VL2 ``true" sky map with a parametrization of the LAT PSF and
calculate the Poisson likelihood $L_{1}$ for the binned Monte-Carlo results.  We
repeat the calculation using a map of a point-source convolved with the PSF and
calculate a second likelihood $L_2$ to arrive at a test statistic that compares
the  subhalo and point-source hypotheses:
\begin{equation}
{\rm TS}=2\ln\left(\frac{L_{1}}{L_{2}}\right).
\end{equation}

The PSF model used for the convolution is a simplification of what is in the
\emph{gtobssim} Monte Carlo.  We assume that there is no azimuthal dependence around the
source direction, while the radial distribution is given by the functional form
used
in the Science Tools to model the non-Gaussian tails of the LAT PSF:
\begin{equation}
p(r)={r\over \delta^2}\left( 1-{1\over\gamma}\right)\left[ 1+ {1\over
2\gamma}\left( {r\over\delta}\right)\right]^{-\gamma}
\label{psf_eq}
\end{equation}
where $\gamma=2$.
The 68\% containment radius of this distribution is about $2.9\delta$.  We match
that radius in each energy bin to the documented LAT 68\% containment
angles~\citep{Atwood2009}, 
taking into account the different PSF for conversions in thick tungsten
versus thin and the relative LAT effective areas for the two conversion types.  
The convolved subhalos should fit well to their simulated counterparts, 
which we verified to be the case for the subhalos of interest.

The TS is only an indicator of how well the data could in principle
statistically distinguish an extended subhalo from a point source.  To rule out
a point-source hypothesis at some confidence level would require something
different, such as a goodness-of-fit test for that hypothesis.  Success would
rely on a thorough Monte-Carlo modeling of the PSF, including a full detector
simulation (\emph{gtobssim} uses only a parametrization of the response), plus
understanding of any associated systematic problems.  Furthermore, this method
of calculating the $TS$ is not applicable to analysis of data, for which the
true subhalo shape is unknown.  Data analysis will require fitting to a subhalo
model with a-priori unknown parameters and will also require dealing with
backgrounds such as partially overlapping point sources, which are not
considered here.

\section{Results}

\begin{figure}[h]
\psfig{file=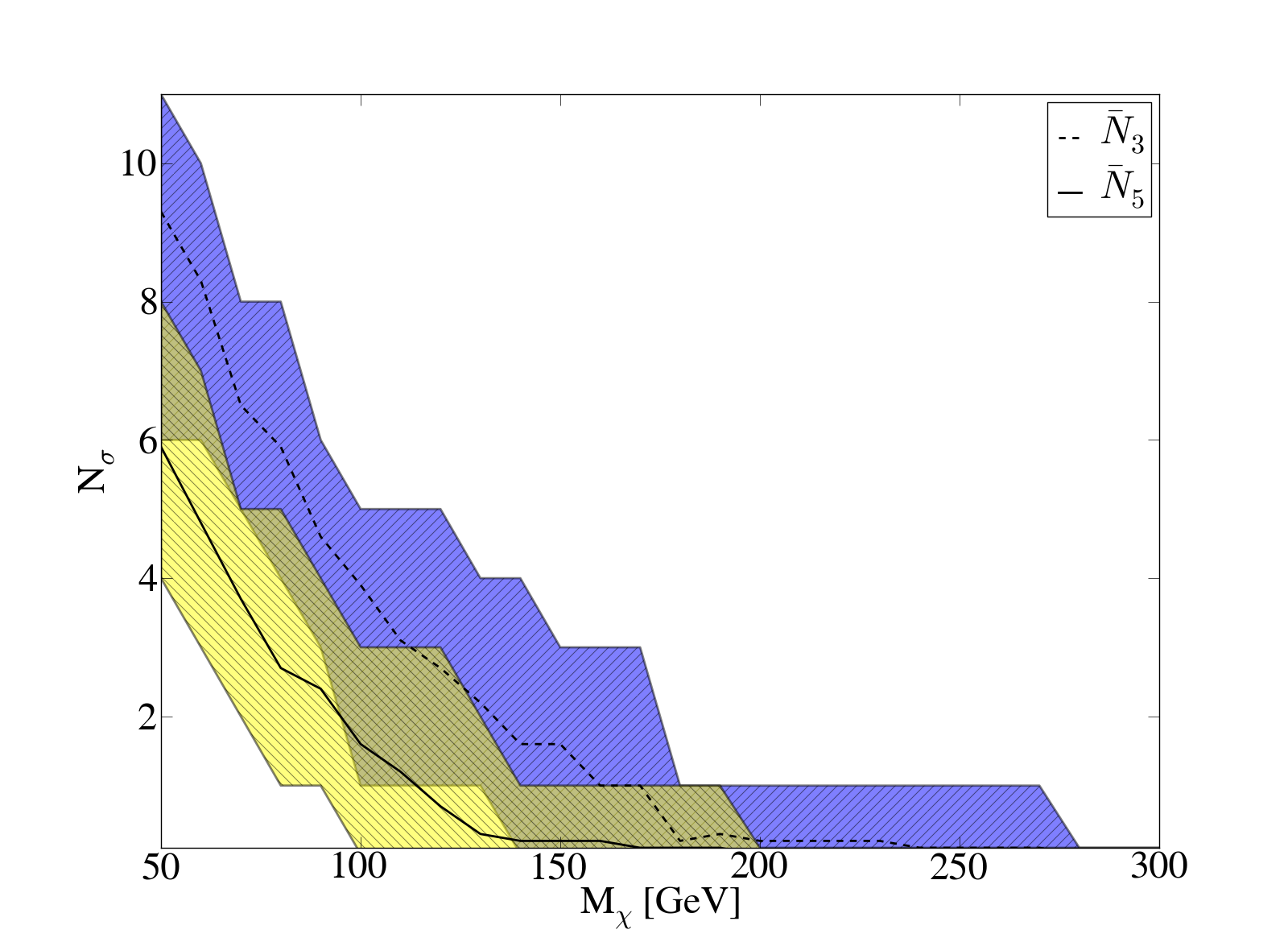,height=2.4in}
\psfig{file=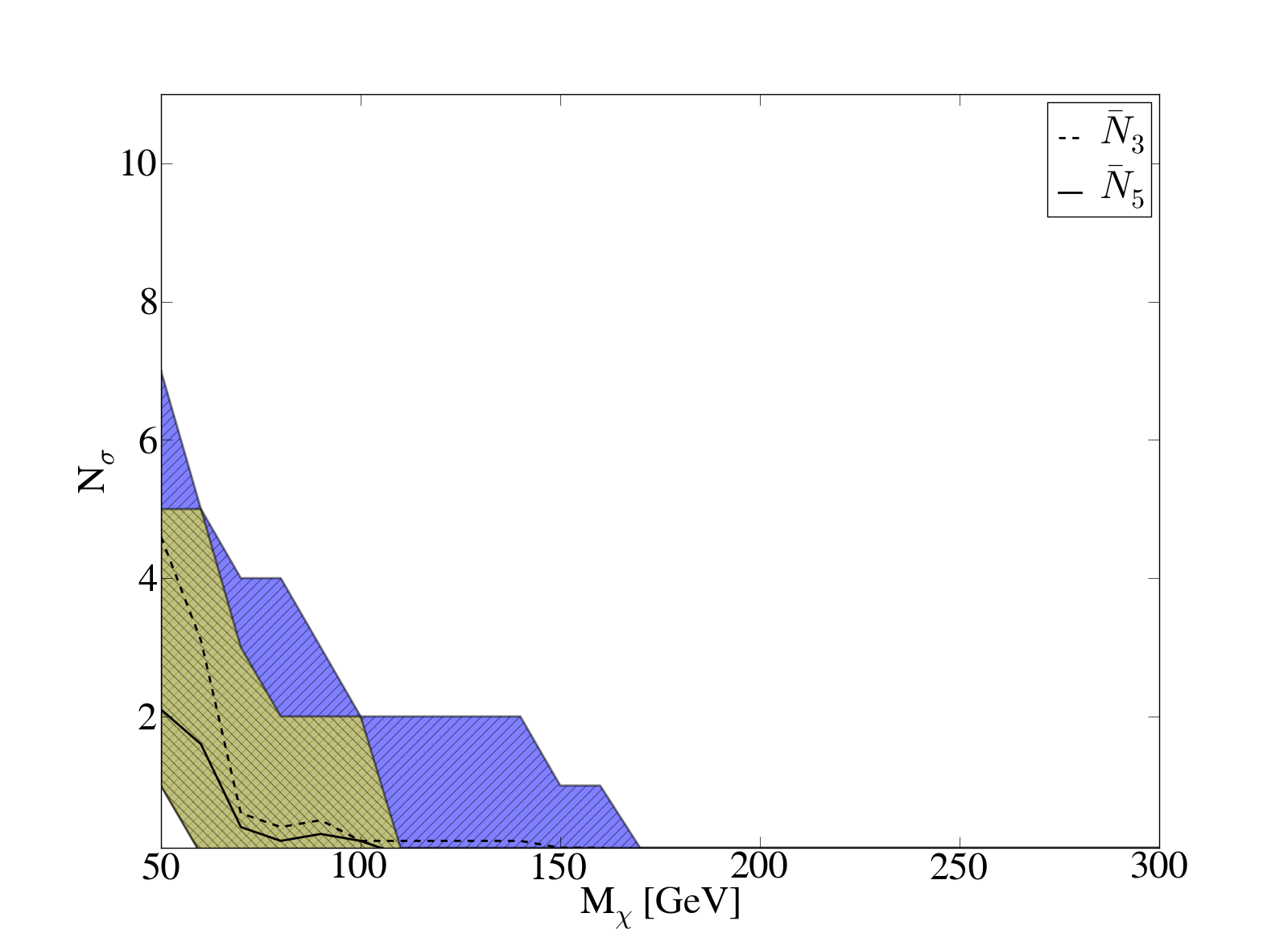,height=2.4in}
\caption{The number of subhalos above five (and three) standard deviations
significance, $N_{5}$ ($N_{3}$), as a function of WIMP mass, $M_{\chi}$,
for $\langle \sigma v\rangle=3\times 10^{-26}{\rm cm}^{3}{\rm s}^{-1}$.  The
shaded regions show the range of variation among ten different observer
positions, while the dashed and solid lines represent the average over all
positions.  The subhalos on the left have been boosted for unresolved
substructure while
those on the right have not.  The simulations represent a
Fermi-LAT observation time of ten years.}
\label{fig_N5}
\end{figure}

Figure~\ref{fig_N5}~shows our results for the number of detectable subhalos with
significance greater than five (and three) versus WIMP mass.  The two
plots represent the different assumptions on the boost of the subhalo
luminosity, as discussed in \S~\ref{sec_methods}.  For each the shaded region
indicates the range of variation among the ten different observer positions
around the Galaxy with a dark line indicating the average.

We find that $N_{5}$ ranges from eight to zero over the range of $M_{\chi}$ and
subhalo boosts.  In all cases, there are no detections above $5\sigma$ with a
WIMP mass greater than 200~GeV.  For comparison, the fiducial model
considered in \citet{Kuhlen2008}, where
$$(M_{\chi}/{\rm GeV},\langle \sigma v\rangle/(10^{-26}{\rm cm}^{3}{\rm
s}^{-1}),\alpha,m_{0}) = (100,3,2.0,10^{-6})\, ,$$
predicted an $N_{5}$ ranging from 13 to 19.  Scaling this down to account
for ten years of orbit, including trigger and SAA dead time, leaves
approximately 11 to 16 $N_{5}$.
For that same DM setup, but with the new, more detailed LAT-specific
analysis and the improved substructure boost implementation, we find here
results for $N_5$ ranging from zero to three with an average over all positions
of 1.6.  This difference comes largely from a lower effective area and angular 
resolution than that assumed in \citet{Kuhlen2008}.  Other factors include the 
useage of Poisson statistics rather than a simplified $S=N_s/\sqrt{N_b}$, 
and a factor of two (upward) error in the \citet{Kuhlen2008} flux calculation.

\begin{figure}[h]
\centerline{\psfig{file=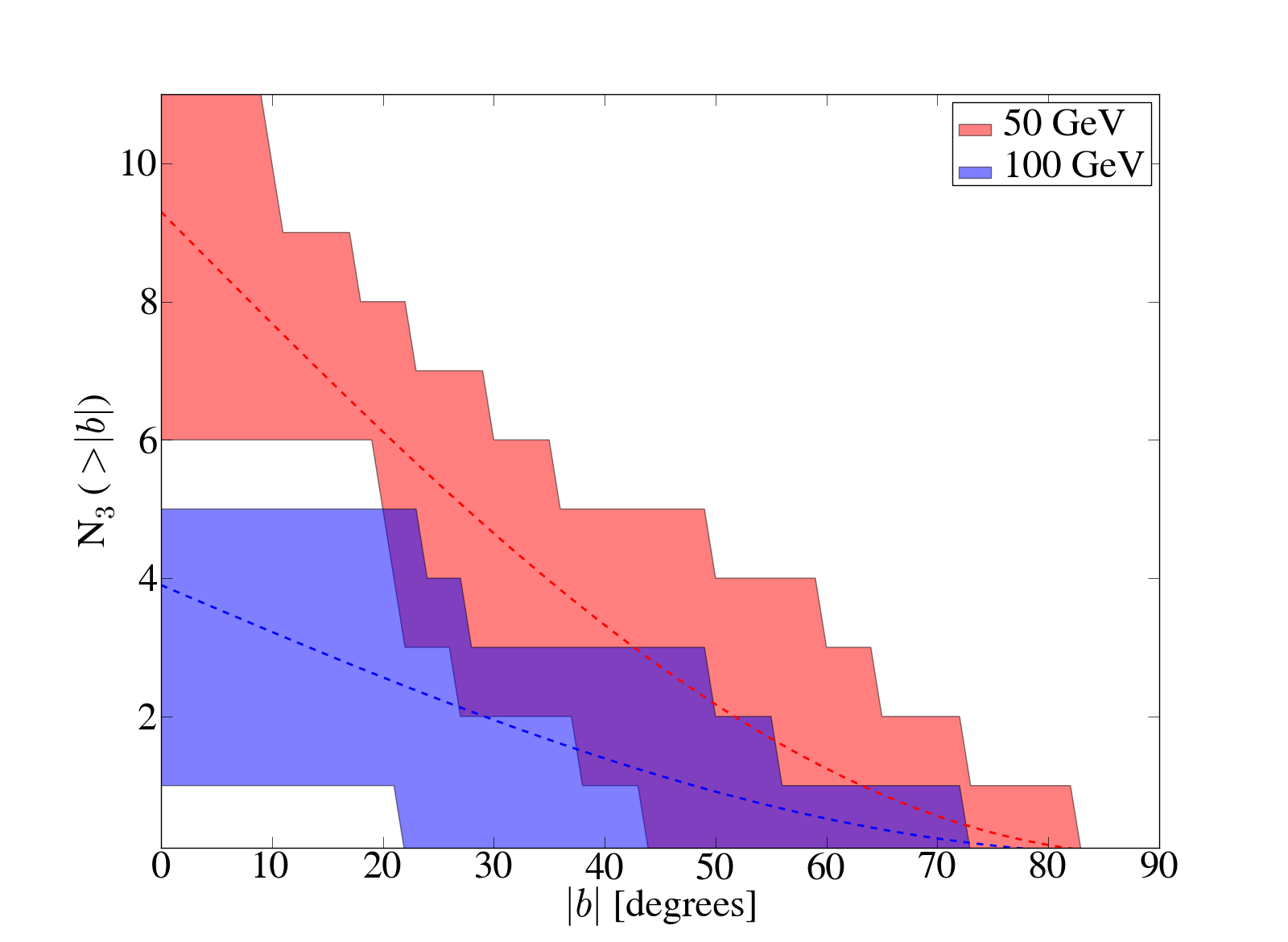,
height=2.8in}
}
\caption{The cumulative distribution of $N_3$ over galactic latitude for two
choices of WIMP mass.  The subhalos here are boosted and the shaded regions
span the variations due to different observer positions around the Galaxy.  The
dashed lines represent the expected behavior for an isotropic distribution.}
\label{fig_bdist}
\end{figure}

\begin{figure}[h]
\centerline{\psfig{
file=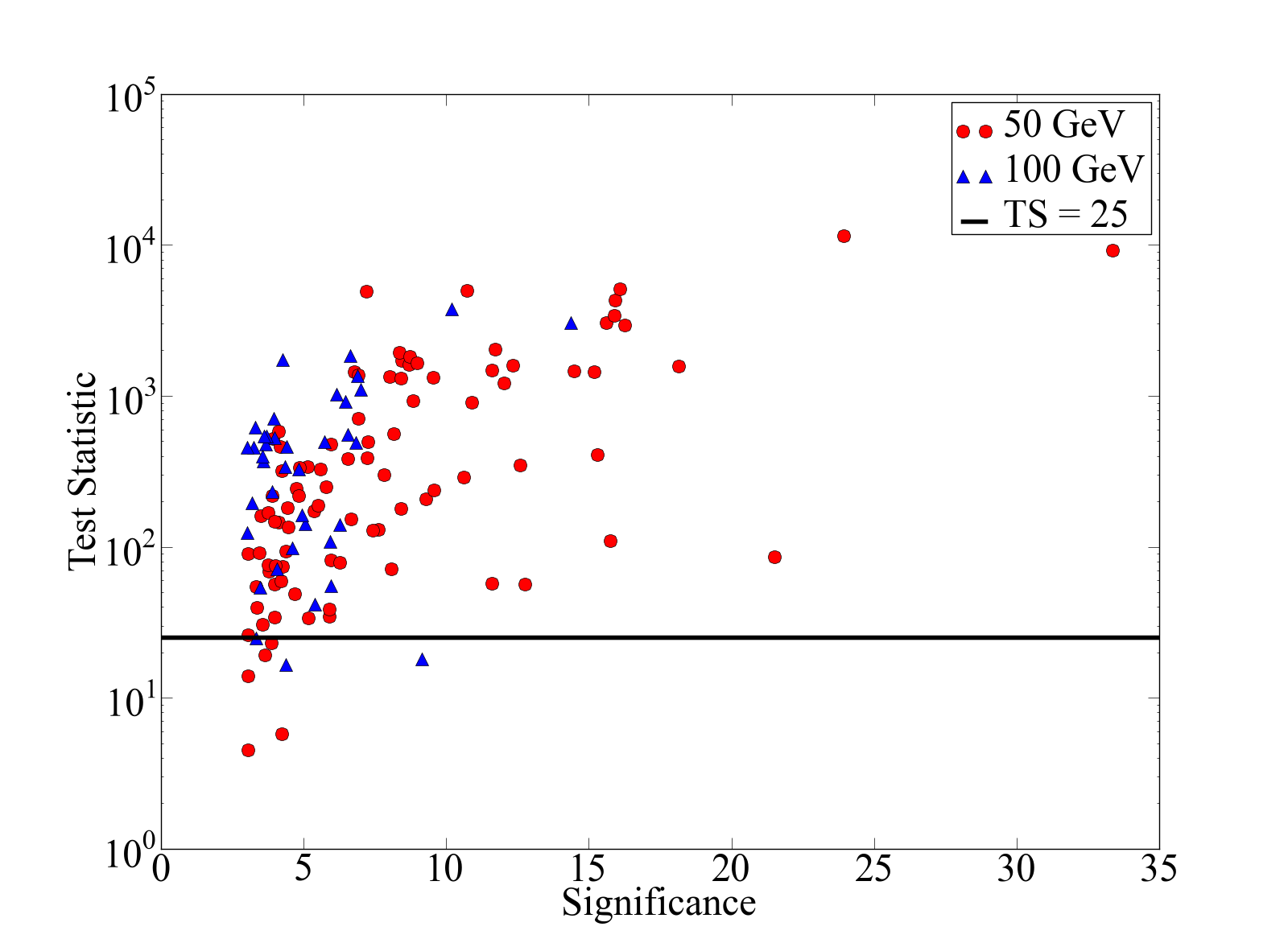,
height=2.8in
}}
\caption{The significance versus test statistic (TS) of $S\geq3$
boosted subhalos, including all Galactic positions.  The
line marks the cut made at ${\rm TS}=25$.  Subhalos
above this line are significantly ($\simeq 5 \sigma$) more extended
than point sources.}
\label{fig_TS_cut}
\end{figure}

These detections are distributed roughly isotropically over the sky, as
can be seen in Fig.~\ref{fig_bdist}.  Requiring that a subhalo have resolvable
angular extent (i.e. that TS $\geq 25$ ($\simeq5\sigma$)), removes five of the
93 $N_{5}$ subhalos from the set of all 50~GeV Galactic positions, and three
of 39 in the 100~GeV set.  Overall, 95\% of $S>3$ detections have $TS>25$. 
This drops to 68\% for the case with no sub-substructure boost.  A 
plot of the relation between TS and significance is given in Fig.~\ref{fig_TS_cut}.

While we plot $N_{5}$ here only as a function of $M_{\chi}$, the
signal strength is also dependent on the annihilation
cross section, $\langle \sigma v\rangle$, a quantity that has a plausible
range spanning orders of magnitude.  
To a lesser extent,
uncertainty in the the boost parameters $\alpha$ and $m_{0}$ and
the nature of the central extrapolation of the density profile,
i.e. whether an NFW or Einasto profile is assumed, can all make roughly factors
$\sim 2$ differences in the $N_{5}$ ~\citep{Kuhlen2008}.

\section{Discussion}

We have estimated the sensitivity of the Fermi-LAT to WIMP annhilation
gamma-rays from Galactic DM substructure predicted by the VL2
simulation.  Using a thorough instrument simulation, we add to predictions
such as Kuhlen et al. 2008 a more accurate treatment of the LAT's
capabilities.  Many of the details of the instrument's behavior, e.g.\ reduced
live-time due to the SAA, a different PSF in the thick and thin foil converters,
etc. lead to a less optimistic predicted sensitivity.  Combined with a less
concentrated sub-substructure boost, these factors leave room for very few
expected detectable subhalos over the LAT's lifetime, given a conventional DM
candidate.

Beyond the pure signal-to-noise ratio, resolution of the angular extension of a
subhalo would go a long way to make it a convincing direct DM
detection candidate.  By computing a test-statistic for each possible predicted
detection, we estimate the fraction which could be confidently deemed extended.
Roughly 83\% of our significant detections should have $\geq5\sigma$ resolution
of angular extent.  This is a somewhat optimistic prediction as we use the true
subhalo density profile in our likelihood ratio, instead of a generic (NFW,
Einasto, etc..)  profile as would probably be used in a real blind substructure
search.  Even so, based on this study of subhalos resolved in the VL2 simulation,
it seems likely that a candidate bright enough to be significant will also appear 
extended.

If the Fermi-LAT does not resolve any significant DM substructure, flux limits
will be of value.  In this case it is important to understand the astrophysical
uncertainties in predictions like the one given here.  The strongest of these
come from the extrapolations needed to compensate for finite numerical
resolution in N-body simulations such as VL2, and the stochastic nature of our
position in the Galaxy relative to individual subhalos.  Simulating signals from
multiple random positions at our Galactic radius gives an estimate of the latter
uncertainty.  We account for unresolved sub-subhalos within subhalos
resolved in VL2 by the boost correction Eqn. (1). Comparison with
un-boosted results illustrate that this effect is of some importance,
while its exact amplitude remains uncertain. As in Kuhlen et
al (2008 and 2009) we have neglected any possibly significant signals
from subhalos that are too small to be resolved in the VL2
simulation and we refer to these works for detailed discussions of
their potential detectability:  \citet{2008MNRAS.384.1627P,2008JCAP...10..040S,2009PhRvD..80b3520A}.

Note also that in models in which the pair annihilation of
non-relativistic DM particles is enhanced by the Sommerfeld effect
\citep[e.g.][]{Arkani-Hamed2009}, the subhalo signal can be strongly
enhanced owing to the lower velocity dispersion of the DM in subhalos
\citep{Robertson2009,Lattanzi2009}. Even the first year of Fermi-LAT data
will strongly constrain many such scenarios \citep{Kuhlen2009b,Bovy2009,Kistler2009}.

We look forward to the Fermi team's upcoming analyses of the first
year's LAT data, searching for a DM annihilation signal from dark subhalos.  An 
analysis of the DM signal from known dwarf galaxies is already published \citep{AbdoDMSat}.

\acknowledgments Support for this work was provided by NASA through
grants HST-AR-11268.01-A1, NNX08AV68G (P.M.), and NASA Guest Investigator grant
NNX08AV59G (B.A.).  This research used resources of the National Center for
Computational Sciences at Oak Ridge National Laboratory, which is
supported by the Office of Science of the Department of Energy under
Contract DE-AC05-00OR22725.
Fermi-LAT simulation tools were provided by the Fermi-LAT collaboration and the
Fermi Science Support Center.

{}

\newpage


\end{document}